\documentclass{article}

\usepackage[preprint]{neurips_2025}

\usepackage[T1]{fontenc}    
\usepackage{hyperref}       
\usepackage{url}            
\usepackage{booktabs}       
\usepackage{amsfonts}       
\usepackage{nicefrac}       
\usepackage{microtype}      
\usepackage{xcolor}         
\usepackage{cite}
\usepackage{amsmath,amssymb,amsfonts}
\usepackage{algorithmic}
\usepackage{graphicx}
\usepackage{textcomp}
\usepackage{xcolor}
\usepackage{url} 
\usepackage[utf8]{inputenc} 
\usepackage{booktabs} 
\usepackage{graphicx}
\usepackage{graphicx}
\usepackage{subcaption}
\usepackage{amsmath}  
\usepackage{amssymb}  
\usepackage{amsthm}  
\usepackage{comment}

\title{TRUST – Transformer-Driven U-Net for Sparse Target Recovery}

\author{%
  Di An\textsuperscript{1} \quad
  Dylan Poppert\textsuperscript{1} \quad  
  Jiayue Li\textsuperscript{1} \quad
  Mark Foster\textsuperscript{1} \quad
  Trac D. Tran\textsuperscript{1} \\
  \textsuperscript{1}Department of Electrical and Computer Engineering, Johns Hopkins University \\
  \texttt{\{dan5, dpopper2, jli275, mark.foster, trac\}@jhu.edu}
}

\date{} 

\begin{document}

\maketitle

\begin{abstract}

In the context of inverse problems $\bf y = Ax$, sparse recovery offers a powerful paradigm shift by enabling the stable solution of ill-posed or underdetermined systems through the exploitation of structure, particularly sparsity. Sparse regularization techniques via $\ell_0$- or $\ell_1$-norm minimization encourage solutions $\bf x$ that are both consistent with observations $\bf y$ and parsimonious in representation, often yielding physically meaningful interpretations. In this work, we address the classical inverse problem under the challenging condition where the sensing operator $\bf A$ is unknown and only a limited set of observation-target pairs $\{ \bf x,\bf y \}$ is available. We propose a novel neural architecture, TRUST, that integrates the attention mechanism of Transformers with the decoder pathway of a UNet to simultaneously learn the sensing operator and reconstruct the sparse signal. The TRUST model incorporates a Transformer-based encoding branch to capture long-range dependencies and estimate sparse support, which then guides a U-Net-style decoder to refine reconstruction through multiscale feature integration. The skip connections between the transformer stages and the decoder not only enhance image quality but also enable the decoder to access image features at different levels of abstraction. 
This hybrid architecture enables more accurate and robust recovery by combining global context with local details. Experimental results demonstrate that TRUST significantly outperforms traditional sparse recovery methods and standalone U-Net models, achieving superior performance in SSIM and PSNR metrics while effectively suppressing hallucination artifacts that commonly plague deep learning-based inverse solvers.

\end{abstract}

\section{Introduction}


The linear inverse problem is fundamental to modern signal processing, statistical modeling, and machine learning. The typical model here is $\bf y = A x + w$, where we seek to recover an unknown signal ${\bf x} \in \mathbb{R}^n$ from a set of potentially noisy measurements ${\bf y} \in \mathbb{R}^m$ using the sensing matrix or the sensing operator ${\bf A} \in \mathbb{R}^{m \times n}$. This problem arises in a wide range of scientific and engineering applications, including magnetic resonance imaging (MRI), computed tomography (CT), optical imaging, geophysics, astronomy and remote sensing, where observations are often limited, incomplete, noisy or partially corrupted \citep{Tibshirani, InverseProblemComp, InverseProblem, LinearInverse}.


Classical approaches to solving inverse problems have been significantly advanced by the theory of compressed sensing (CS) and associated sparse recovery methods \citep{candes2006stable, donoho2006compressed, candes2006robust, elad2010sparse}. 
These techniques leverage the fact that many natural signals are sparse or compressible in specific transform domains, such as wavelets, gradients, or learned dictionaries. Under suitable conditions on the sensing matrix $\bf A$, CS guarantees accurate recovery of sparse signals from far fewer measurements than traditionally required. The reconstruction problem is typically posed as follows
\begin{equation}
\min_x \| {\bf x} \|_0 \quad \text{subject to} \quad \| {\bf A x - y} \|_2 \leq \epsilon
\quad \text{or} \quad 
    \min_x \| {\bf x} \|_1 \quad \text{subject to} \quad \| {\bf A x - y} \|_2 \leq \epsilon
\label{eq:l0l1}
\end{equation}
\noindent where the $\ell_0$- or $\ell_1$-norm promotes sparsity in $\bf x$ and the constraint enforces fidelity to the measurements $\bf y$. While these methods are mathematically principled and offer performance guarantees, they rely on accurate knowledge of the sensing operator
$\bf A$ and assume linearity -- assumptions that often break down in more complex or nonlinear measurement settings.


Deep learning has recently emerged as a powerful data-driven alternative to mitigate the limitations of classical approaches. In particular, convolutional neural networks (CNNs), notably encoder-decoder architectures like U-Net \citep{Ronneberger2015UNet} have shown strong performance in tasks such as denoising \citep{zhang2017beyond, zhang2018ffdnet}, super-resolution \citep{ledig2017photo} and compressive image recovery \citep{mousavi2015deep}. 
These models learn to map raw sensor measurements directly to reconstructed signals, promising end-to-end inverse modeling, eliminating the need for hand-crafted priors, and enabling greater adaptability to real-world variations. This is particularly impactful in domains like synthetic aperture radar (SAR) and computational optics, where the forward process involves nonlinear physics such as diffraction or phase retrieval that are analytically intractable \citep{rivenson2018phase, jin2017deep}. 
These methods not only improve reconstruction quality, but also generalize well when trained on realistic measurement-target pairs.




Despite these advances, cross-domain inverse problems—where measurement and target domains are fundamentally different—remain a substantial challenge. For example, in optical systems, the relationship between observations and desired reconstructions is often nonlinear and ambiguous. Additionally, standard CNNs are inherently limited by their local receptive fields and spatial inductive biases, making it difficult to capture the global context and long-range dependencies essential for resolving such ambiguities. To overcome these limitations, researchers have begun exploring transformer-based architectures, which leverage self-attention mechanisms to model global interactions across spatial regions \citep{dosovitskiy2020image, chen2021transunet}.
These models have shown remarkable success in high-level vision tasks and are increasingly being adopted in low-level inverse problems.

In this work, we introduce a novel architecture called TRUST, a transformer-driven U-Net for sparse target recovery that integrates the Vision Transformer (ViT) with U-Net for optical image reconstruction. Unlike only convolution blocks that primarily rely on local filtering, the attention mechanism successfully captures global dependencies across image patches, making them especially suited for cross-domain reconstruction tasks. Extensive experiments demonstrate that TRUST consistently outperforms traditional compressed sensing methods and state-of-the-art deep learning models.

\section{Problem Definition}


In this paper, we address the classical inverse problem $\mathbf{y} = A \mathbf{x} + \mathbf{w}$ via sparse recovery as in (\ref{eq:l0l1}) under the challenging condition where \textit{the sensing operator} $\mathbf{A}$ \textit{is unknown} and \textit{we only have access to a limited set of available observation-target pairs $\{ \mathbf{x}, \mathbf{y} \}$ as training data}. Note that both the measured data $\mathbf{y}$ and the target images $\mathbf{x}$ are commonly flattened into vectors for mathematical convenience, although they originally represent structured two-dimensional spatial information. 





Solving this ill-posed inverse problem using classical sparsity-driven methods would typically require first approximating the unknown operator $\mathbf{A}$ via dictionary learning techniques \citep{Aharon2006KSVDSparse}, followed by applying sparse recovery algorithms such as Orthogonal Matching Pursuit (OMP) \citep{Tropp2007OMP} or the Fast Iterative Shrinkage-Thresholding Algorithm (FISTA) \citep{Beck2009FISTA}. However, this two-step approach is often inefficient, particularly in complex or nonlinear sensing environments \citep{Tarantola2005Inverse, Vogel2002Inverse}. As an alternative, we adopt modern deep learning-based strategies, specifically U-Net \citep{Ronneberger2015UNet} and the proposed TRUST architecture, which directly learn the inverse mapping from data. These models eliminate the need for explicit knowledge of the sensing matrix while simultaneously enabling accurate reconstruction of sparse target signals \citep{Mardani2019DeepMRI}.  

Throughout this paper, we motivate the development of the proposed TRUST network and illustrate its working concept in the context of a practical noninvasive coded aperture multicore fiber microendoscope for brain imaging \citep{Willett2007CodedAperture, Farahi2013DeepTissue}, capable of capturing sub-micron spatial image features.  

\begin{figure}[tb]
\centering
\includegraphics[width=\linewidth]{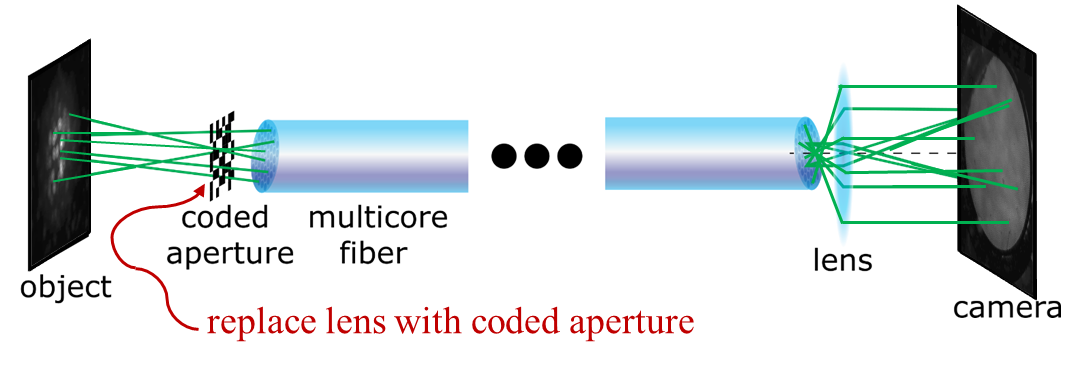} 
\caption{A multicore fiber coded aperture microendoscope. The fiber bundle contains around 6000 cores, has a diameter of 270 $\mu m$, capable of capturing sub-micron image features.}
\label{model:CodedAperture}
\end{figure}


\section{TRUST} \label{sec:criteria}

\subsection{Previous Works}

Numerous efforts have been made to address the sparse recovery problem using deep learning. Early pioneering approaches, such as ISTA-Net \citep{ISTA-Net} and ADMM-Net \citep{ADMM-Net}, belong to the class of algorithm unrolling methods \citep{Unrolling}. These architectures translate each iteration of a classical sparse optimization algorithm into a corresponding layer of a neural network, allowing the model to learn key parameters while preserving the interpretability of the original iterative structure. Although unrolling networks offer advantages in terms of interpretability, parameter efficiency, and performance in structured or low-data regimes, they generally fall short when applied to large-scale complex recovery tasks.

In contrast, more general-purpose architectures like U-Net have emerged as dominant solutions in signal and image reconstruction. Originally designed for biomedical image segmentation, U-Net’s encoder–decoder structure with skip connections allows it to effectively capture and integrate multiscale features, making it well-suited for complex spatial reconstruction tasks \citep{ronneberger2015u}. Recent advancements such as TransUNet \citep{chen2021transunet} further enhance U-Net’s capabilities by incorporating attention mechanisms at the network bottleneck, leveraging the strength of self-attention to model long-range dependencies and improve global context modeling. In the opposite direction is the fully transformer-based encoder–decoder Restormer \citep{Restomer}, which integrates attention mechanisms with multiscale architectures for image reconstruction.

A closer examination of the linear inverse problem $\bf y = A x$ reveals a fundamental challenge: {\it local features in the signal $\bf x$ may become dispersed or diffused across the global observation $\bf y$.} This is particularly true in compressed sensing, where measurements are often acquired in incoherent or randomized domains to satisfy theoretical recovery guarantees. In such settings, reconstruction architectures that primarily rely on local receptive fields—such as classical CNNs or even U-Net—can struggle to recover globally consistent structure, especially when long-range dependencies are critical to disambiguate spatial information.

\subsection{Proposed Architecture}

\begin{figure}[!htb]
\centering
\includegraphics[width=\linewidth]{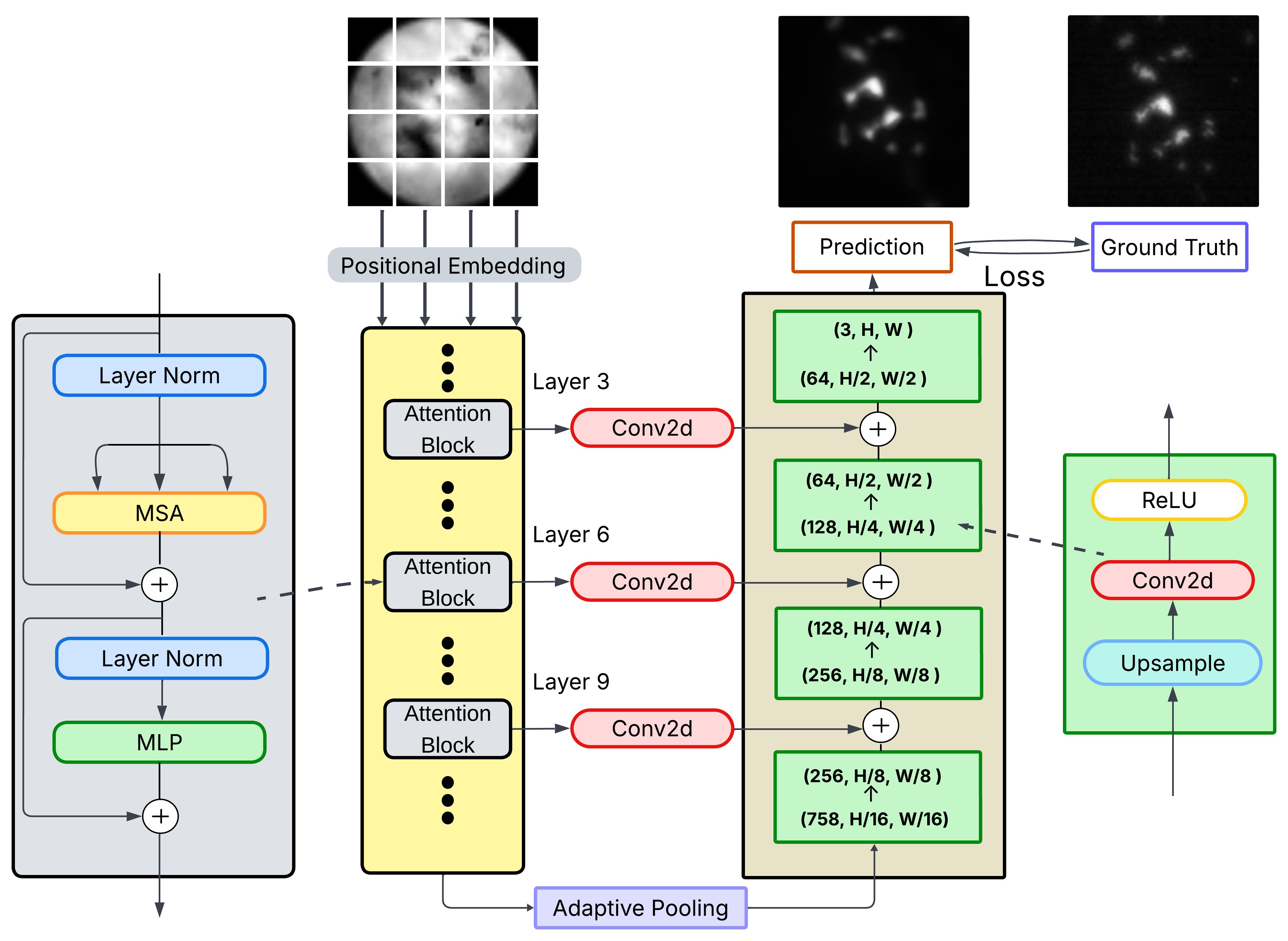} 
\caption{TRUST Architecture -- Transformer-Driven U-Net for Sparse Target Recovery}
\label{fig:model}
\end{figure}

Motivated by these limitations, we propose TRUST, a hybrid architecture designed to combine the strengths of both local and global modeling paradigms. As illustrated in Figure~\ref{fig:model}, TRUST employs a Vision Transformer (ViT) to extract multiscale global attention features from the input, effectively modeling long-range dependencies across the spatial domain. These features are then processed through an adaptive pooling layer, which performs pixel-wise smoothing to enhance robustness and feature continuity. Finally, a U-Net-inspired upsampling pathway incrementally refines the output, progressively recovering fine spatial detail and enforcing structural coherence.

In the remainder of this section, we delve into the design rationale behind each component of the TRUST architecture. We aim to provide a deeper understanding of their individual contributions and their synergistic effect on the network’s overall performance in sparse recovery tasks.



\subsection{Attention Can Be an Excellent Encoder}


Compared to traditional convolutional operations, the attention mechanism in Transformers offers a significant advantage in modeling global contextual relationships across spatial features. At the heart of this mechanism is the self-attention operation, defined as:
\begin{equation}
    \text{Attention}({\bf Q, K, V}) = \text{softmax}\left(\frac{{\bf QK}^T}{\sqrt{d_k}}\right) {\bf V}
    \label{attention}
\end{equation}
\noindent where $\bf Q$, $\bf K$, and $\bf V$ denote the query, key, and value matrices, respectively, and $d_k$ is the dimensionality of the key vectors. This formulation effectively performs a scaled dot-product similarity -- akin to a normalized cosine similarity -- which allows the model to dynamically focus on salient regions and capture long-range structural dependencies across the entire image.
  
We further demonstrate that self-attention applied directly to the measurement domain $\bf y$ can approximate the attention features of the ground truth signal $\bf x$, provided that the sensing matrix satisfies the Restricted Isometry Property (RIP) \citep{RIP}. Specifically, if $\bf A$ satisfies the Restricted Isometry Property (RIP) of order \( 2k \) with RIP constant \( \delta_{2k} \in (0, 1) \), then for all \( 2k \)-sparse vectors \( {\bf z} \in \mathbb{R}^n \), we have
\[
(1 - \delta_{2k}) \; \|{\bf z}\|_2^2 \; \leq \; \|{\bf A z}\|_2^2 \; \leq \; (1 + \delta_{2k}) \; \|{\bf z}\|_2^2.
\]
This implies that the geometry of sparse vectors is approximately preserved under the mapping $\bf A$.
More precisely, the attention error between two representations in two different domains is bounded by the RIP constant as follows (see the Appendix for the detailed derivation):
\[
\left| {\bf y}^\top {\bf y}' - {\bf x}^\top {\bf x}' \right| = \left| {\bf x}^\top {\bf A}^\top {\bf A} {\bf x}' - {\bf x}^\top {\bf x}' \right| \; \leq \;  \delta_{2k}.
\]

\begin{figure}[!ht]
\centering


\centering
\includegraphics[width=0.8\linewidth]{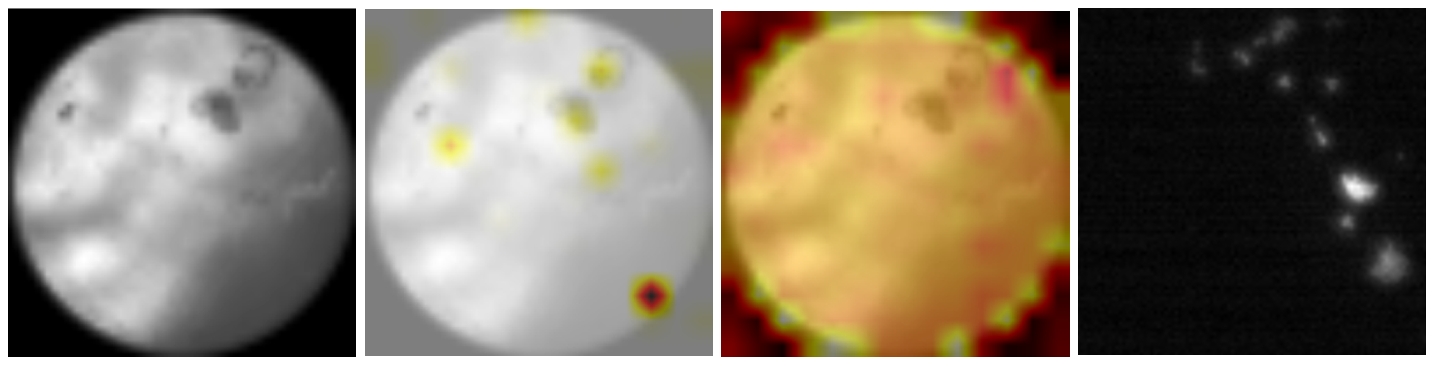} 
\caption{Overlaying attention map of a sample collected from the microendoscope in Figure~\ref{model:CodedAperture}. From left to right: response $\bf y$, single head attention, aggregated multihead attention, and ground truth $\bf x$.}
\label{fig:attention}

\end{figure}

As depicted in Figure~\ref{fig:attention}, the attention map generated from $\bf y$ indeed highlights key spatial structures and regions that closely resemble those in the original image $\bf x$. This empirical observation aligns with our theoretical analysis and confirms that the attention module not only facilitates contextual reasoning, but also plays a critical role in sparse support recovery. These extracted attention features serve as a powerful prior, guiding the subsequent reconstruction stages in our TRUST framework to focus on the most informative regions of the measurement.

\subsection{Adaptive Pooling Layer}

Processing high-dimensional attention feature maps directly for image reconstruction can be both computationally intensive and inefficient in terms of capturing spatial hierarchies. To address this, we introduce an adaptive pooling layer, which serves two critical functions: dimensionality reduction and feature standardization. First, adaptive pooling reduces the spatial dimensions of the attention output, thereby lowering computational cost and enabling the model to focus on semantically meaningful features at a coarser resolution. Second, it ensures that the resulting feature maps are standardized to a fixed output size, regardless of the original input dimensions, maintaining architectural consistency across inputs of varying shapes and sizes \citep{he2015spatial}.

\begin{figure}[!ht]
\centering
\begin{subfigure}{0.4\columnwidth}
\centering
\includegraphics[width=\linewidth]{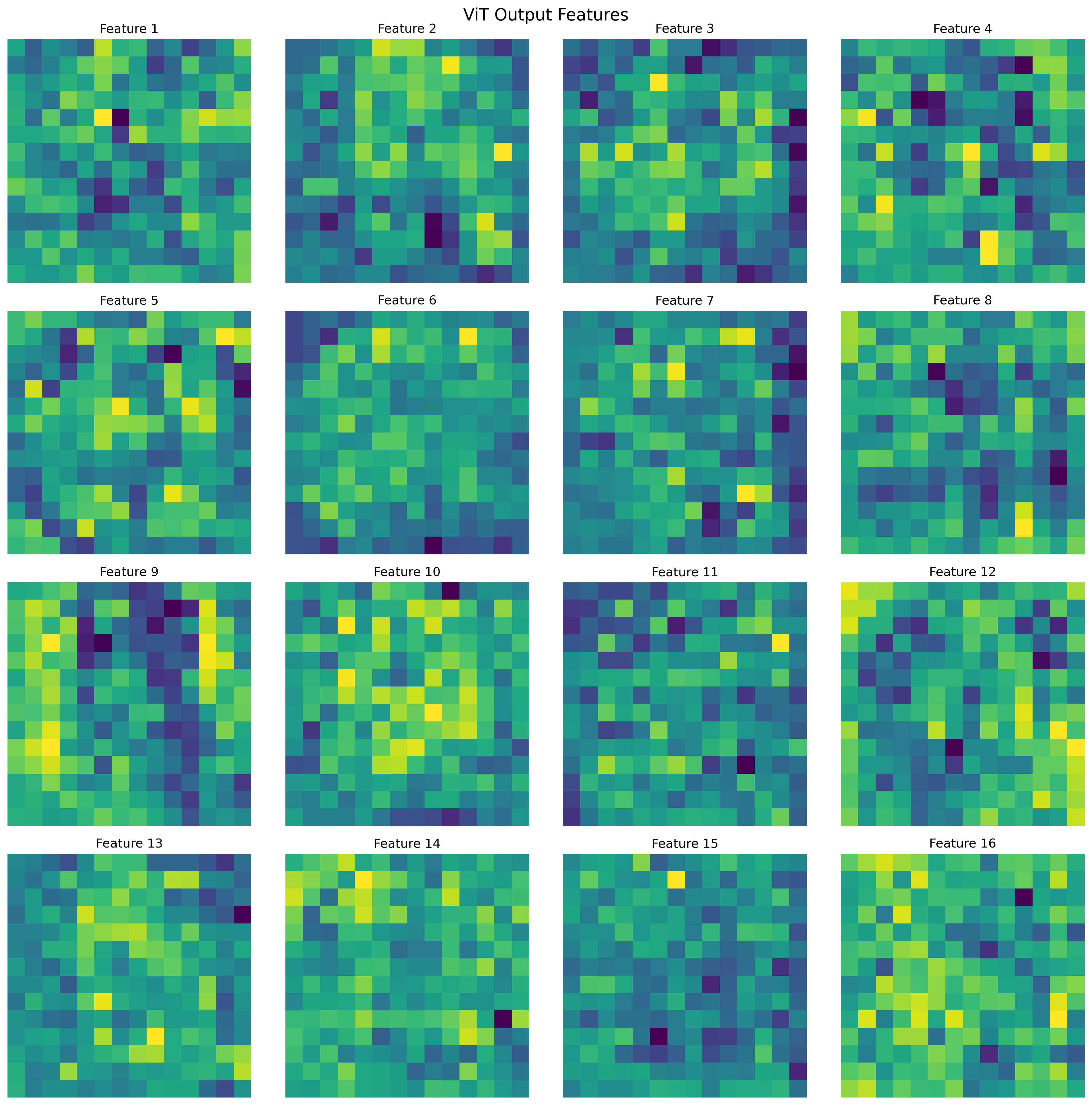} 
\caption{ViT Feature Map}
\label{fig:sim1}
\end{subfigure}%
\hspace{0.02\columnwidth}%
\begin{subfigure}{0.4\columnwidth}
\centering
\includegraphics[width=\linewidth]{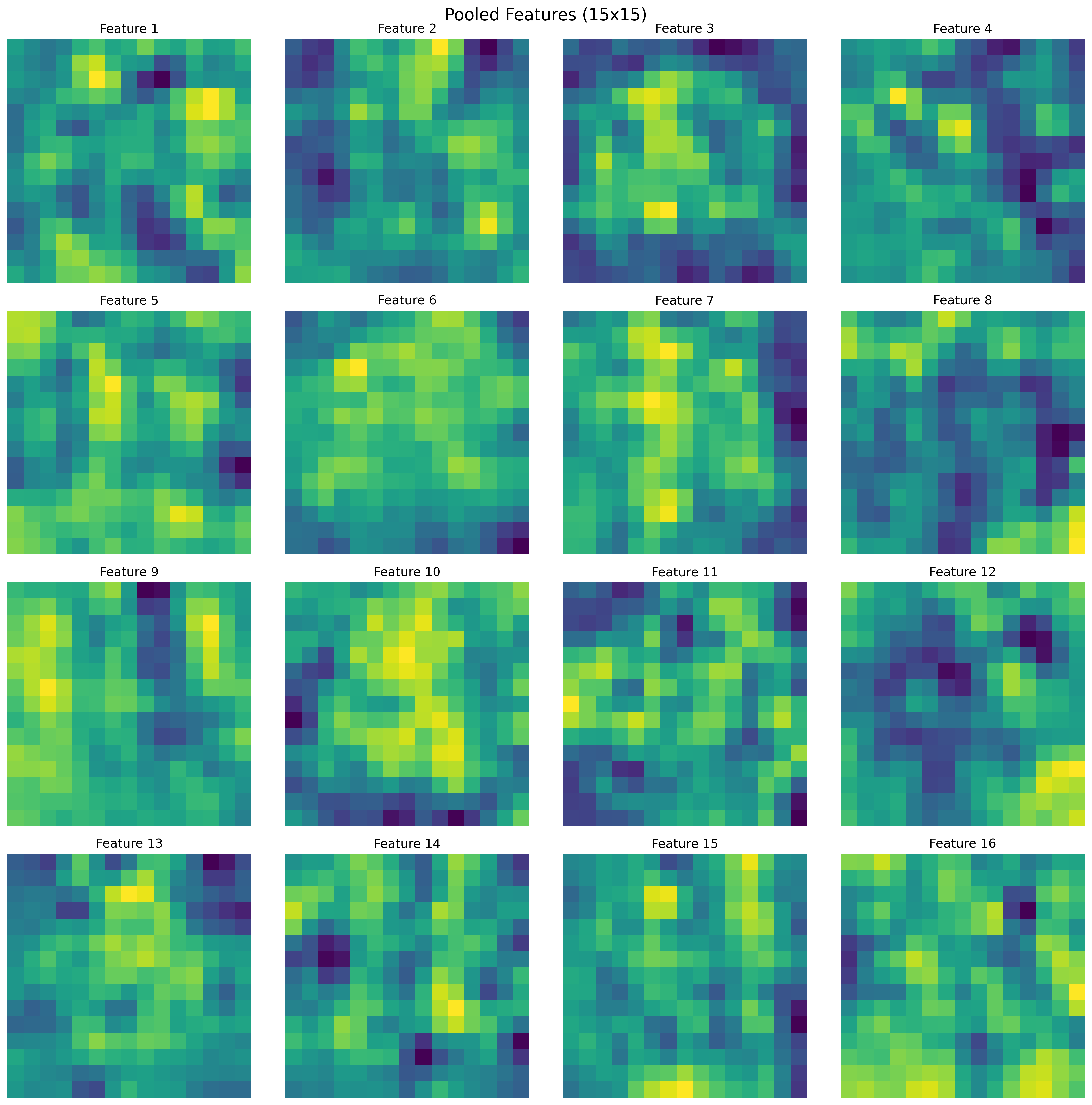}
\caption{Feature Map after Pooling}
\label{fig:sim2}
\end{subfigure}
\caption{Adaptive pooling layer function's effect on a typical attention map.}
\label{fig:adpt pooling}
\end{figure}

As illustrated in Figure~\ref{fig:adpt pooling}, the adaptive pooling layer effectively distills the attention features into a compact representation while preserving their structural integrity. This step is essential for enabling the subsequent decoder stages to reconstruct the image with improved efficiency and precision.



\subsection{U-Net-like Upsampling Decoder for Detail Refinement}
The decoder's primary objective is to reconstruct high-resolution output images from the lower-dimensional feature maps produced by the adaptive pooling layer. To achieve this, our decoder adopts a U-Net-like architecture that progressively restores spatial resolution while refining structural detail. At each decoding stage, the upsampling operation enlarges the feature map dimensions, followed by convolutional layers (Conv2D) that refine spatial content and ReLU activations that introduce nonlinearity and expressive capacity. This sequential refinement pipeline enables the model to recover fine-grained features that may have been compressed or diffused in earlier encoding stages.

\begin{figure}[!ht]
\centering

\centering
\includegraphics[width=\linewidth]{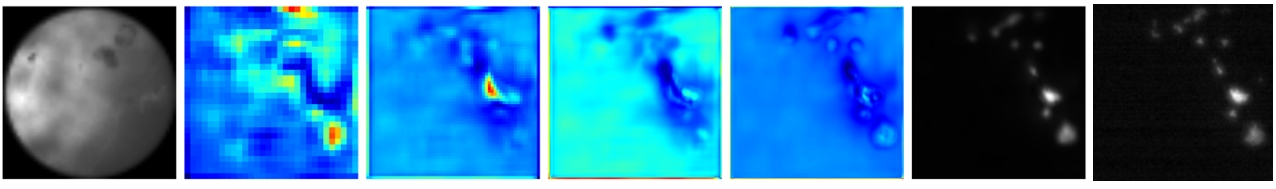} 
\caption{Different stages of decoding. From left to right: response $\bf y$, stage $1$, stage $2$, stage $3$, stage $4$, reconstructed image $\hat{\bf x}$, and ground truth $\bf x$. Resolution is enhanced gradually from left to right.}
\label{fig:decoder_vis}

\end{figure}
As shown in Figure~\ref{fig:decoder_vis}, we visualize the transformation of feature maps through the decoder layers. The initial image, representing the raw diffraction pattern, undergoes a series of attention-driven and convolutional transformations that progressively reveal meaningful structure. At the first decoding stage (30×30×256), the network extracts foundational high-frequency components, with key activations highlighted in red and yellow. As decoding continues through intermediate layers (60×60×128, 120×120×64, and 240×240×32), the feature maps increase in spatial resolution and decrease in channel dimensionality, reflecting a systematic reassembly of the signal’s spatial hierarchy.

This visualization illustrates how our architecture bridges the domain gap between the incoherent diffused measurements and the target image space. The Transformer module captures global long-range dependencies early in the pipeline, while the U-Net decoder gradually reconstructs the local structure through multiscale upsampling and refinement. The evolution of the activation maps shows that the network selectively amplifies salient features and suppresses irrelevant noise, ultimately producing a high-fidelity optical reconstruction. This combination of global context modeling and localized detail recovery is essential for achieving robust and precise image reconstruction in complex sparse inverse recovery tasks.


\section{Experimental Results}

We leverage transfer learning on our proposed TRUST architecture by incorporating the pretrained 'google/vit-base-patch16-224' Vision Transformer as the encoder backbone \citep{dosovitskiy2020image}. This strategic choice significantly accelerates training convergence and improves performance for the specialized task of optical image reconstruction. Training was conducted on a setup with four Tesla P400 GPUs (24 GB VRAM each), using a learning rate of $1 \times 10^{-4}$ and a batch size of $128$. Given the modest computational resources, training was extended over the course of one week to ensure stable convergence and optimal reconstruction quality.

\subsection{Datasets and Evaluation Metrics}
We evaluated TRUST on two datasets: a custom optical imaging dataset obtained from the multicore fiber microendoscope in Figure~\ref{model:CodedAperture} and the single coil knee dataset in the publicly available FastMRI benchmark. This dual evaluation allows us to assess the model's effectiveness both in domain-specific reconstruction and in common/popular inverse imaging scenarios.

For the optical dataset, training data was obtained from two neuron sample slides, while the test data was collected from a third, unseen sample. The training set consists of 32,000 image pairs (diffraction response and ground truth), and the test set includes 16,000 pairs, all acquired at a consistent depth (object-to-microendoscope tip) distance of 100 microns. This deliberate separation between training and testing sets is essential to validate the model’s ability to generalize beyond memorized patterns and to handle new biological structures under consistent imaging conditions.

To further demonstrate the generalization capability of TRUST, we conducted additional experiments on the FastMRI dataset -- a large-scale benchmark jointly developed by Facebook AI Research and NYU Langone Health for accelerated MRI reconstruction \citep{zbontar2018fastmri}. This task fits the ill-posed inverse problem described in Section 2, where the collected observation comes from an undersampled k-space signal processed through a sparse sampling operator $\bf A$. The degraded image, obtained via inverse Fourier transform (IFFT), contains aliasing artifacts. The goal is to reconstruct a high-quality ground truth image from this undersampled and noisy input \citep{lustig2007sparse}.

We assessed TRUST’s performance using the following evaluation metrics: Mean Squared Error (MSE), Mean Absolute Error (MAE), Peak Signal-to-Noise Ratio (PSNR), Structural Similarity Index Measure (SSIM), and False Discovery Rate (FDR) \citep{wang2004image, gonzalez2002digital} .
These metrics collectively evaluate both low-level pixel-level accuracy and high-level perceptual quality. Detailed definition for each metric along with full details on preprocessing/sampling masks above are provided in the Appendix. 

\subsection{Main Recovery Results}
We first evaluated the performance of our model on the optical imaging dataset, comparing traditional sparse recovery methods with modern deep learning-based approaches. In this experiment, both U-Net and TRUST were trained using a combined loss function consisting of the $\ell_2$ loss and SSIM. All hyperparameters were kept approximately consistent across both models. During training, care was taken to allow each model to converge to a comparable loss level, ensuring a fair performance comparison. The choice of loss function was found to significantly affect reconstruction quality and deserved further discussion in Section~\ref{sec:CostFunctions}. For OMP results, we had to first learn an approximation of $\mathbf{A}$ from the training data prior to conventional sparse recovery \citep{lustig2007sparse}. 

\begin{figure}[!htb]
\centering
\includegraphics[width=\linewidth]{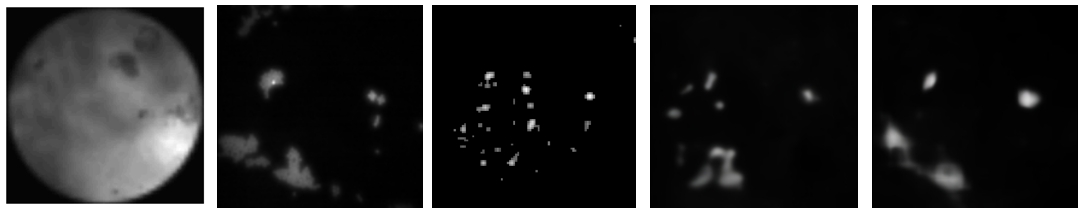} 
\caption{\small Example of reconstruction results with corresponding SSIM and PSNR values. From left to right: response $\bf y$, target $\bf x$, OMP $\{0.301, 68.723dB\}$, U-Net $\{0.779, 71.691dB\}$, TRUST $\{0.862, 72.744dB\}$}.
\label{exp:main_recon_optica}
\end{figure}
\vspace{-0.35in}

\begin{table}[tbh]
\centering
\caption{Average recovery performance on the optics dataset: mean $\pm$ standard deviation}
\label{tab:recon_comparison}
\begin{tabular}{@{}lccccc@{}}
\toprule
\textbf{Method}   & \textbf{MSE}  & \textbf{MAE}  & \textbf{PSNR (dB)}  & \textbf{SSIM}  & \textbf{FDR ($\times$10$^{-2}$)} \\
\midrule
\textbf{OMP}      & 0.0111 $\pm$ 0.0032        & 0.0435 $\pm$ 0.0062        & 68.04 $\pm$ 2.03               & 0.2791 $\pm$ 0.035         & 5.30 $\pm$ 1.03               \\
\textbf{U-Net}    & 0.00451 $\pm$ 0.0022       & 0.0398 $\pm$ 0.012        & 70.76 $\pm$ 2.00               & 0.772 $\pm$0.053          & 1.14 $\pm$ 0.16              \\
\textbf{TRUST}    & \textbf{\textcolor{red}{0.00431 $\pm$ 0.0013}} & \textbf{\textcolor{red}{0.0253 $\pm$ 0.0073}} & \textbf{\textcolor{red}{71.992 $\pm$ 1.94}} & \textbf{\textcolor{red}{0.814 $\pm$ 0.069}} & \textbf{\textcolor{red}{0.901 $\pm$ 0.22}} \\
\bottomrule
\end{tabular}
\end{table}

As demonstrated in Figure~\ref{exp:main_recon_optica} and Table~\ref{tab:recon_comparison}, TRUST outperforms U-Net and classical baselines on a test set of 5,000 randomly selected optical samples. In particular, TRUST produces reconstructions with fewer hallucinations and artifacts, consistent with our theoretical arguments on its ability to leverage global contextual information. This advantage is visually evident in the sample reconstruction where the U-Net prediction exhibits a hallucinated structure near the bottom-left corner, while TRUST successfully suppresses this anomaly and recovers a more faithful representation.

\begin{figure}[!htb]
\centering
\begin{subfigure}{0.49\columnwidth}
\centering
\includegraphics[width=\linewidth]{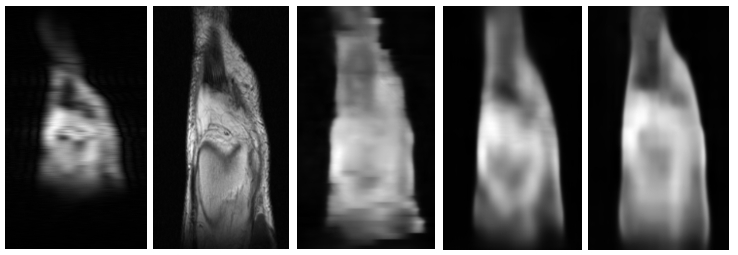} 
\caption{\small SSIM, PSNR (dB): U-Net $\{0.662, 19.03 \}$, Restomer $\{0.638, 19.26\}$, TRUST $\{0.71, 22.00\}$}
\label{fig:mri1}
\end{subfigure}%
\hspace{0.01\columnwidth}%
\begin{subfigure}{0.49\columnwidth}
\centering
\includegraphics[width=\linewidth]{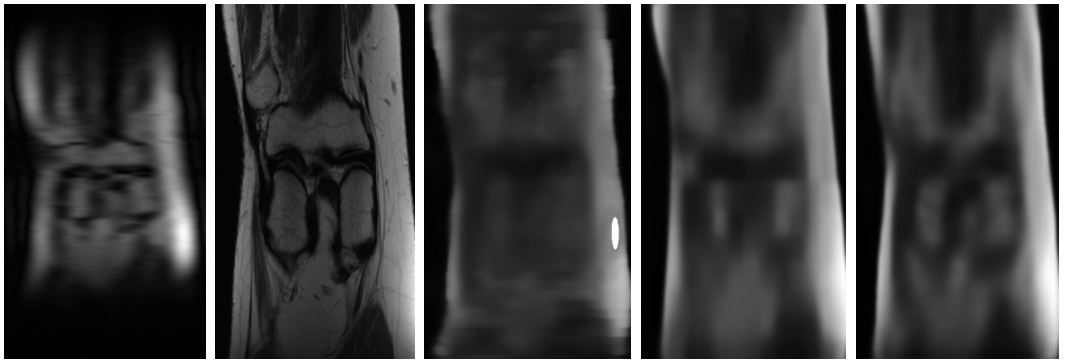}
\caption{\small SSIM, PSNR (dB): U-Net $\{0.631, 20.98\}$, Restomer $\{0.674, 24.25\}$, TRUST $\{0.689, 24.48\}$}
\label{fig:mri2}
\end{subfigure}

\caption{\small Two examples of FastMRI reconstruction results. From left to right: undersampled aliased image, true target $\mathbf{x}$, U-Net reconstruction, Restormer reconstruction, and TRUST reconstruction.}
\label{fig:MRI_res}
\end{figure}

To further evaluate TRUST’s generalizability, we tested its performance on the large-scale standardized FastMRI dataset. Table~\ref{tab:MRI} summarizes the results across 36 randomly selected slices from 108 subjects, totaling approximately 3,000 test images, whereas Figure~\ref{fig:MRI_res} depicts a typical reconstruction sample. These experiments validate that TRUST is not only effective in specialized optical recovery tasks, but also performs competitively in real-world, large-scale medical imaging scenarios \citep{Restomer, ho2020denoising}, showing remarkable adaptability to different domains and reconstruction scenarios. Here, we decide to bypass the underperforming OMP by the more competitive all-transformer Restormer \citep{Restomer}. 



\begin{table}[htb]
\centering
\caption{Average recovery performance on the FastMRI dataset: mean $\pm$ standard of deviation}
\label{tab:MRI}
\begin{tabular}{@{}lccccc@{}}
\toprule
\textbf{Method} & \textbf{MSE} & \textbf{MAE} & \textbf{PSNR (dB)} & \textbf{SSIM} & \textbf{FDR($\times$10$^{-2}$)} \\
\midrule
\textbf{U-Net}  & 0.0861 $\pm$ 0.0246      & 0.0506 $\pm$ 0.0174      & 21.70 $\pm$ 2.74            & 0.668 $\pm$ 0.0900         & 4.26 $\pm$ 4.99       \\
\textbf{Restormer}  & 0.0692 $\pm$ 0.0227      & 0.0411 $\pm$ 0.0160       & 23.72 $\pm$ 3.15              & 0.698 $\pm$ 0.0953        & 2.97 $\pm$ 4.74       \\
\textbf{TRUST}  & \textbf{\textcolor{red}{0.0613 $\pm$ 0.0220}} & \textbf{\textcolor{red}{0.0353 $\pm$ 0.0133}} & \textbf{\textcolor{red}{24.81 $\pm$ 3.13}} & \textbf{\textcolor{red}{0.717 $\pm$ 0.0851}} & \textbf{\textcolor{red}{2.78 $\pm$ 4.33}} \\
\bottomrule
\end{tabular}
\end{table}

\captionsetup[sub]{font=small, justification=centering, singlelinecheck=true}

\subsection{Ablation Study: Impact of Loss Function Choice}
\label{sec:CostFunctions}

\begin{figure}[!ht]

\centering
\includegraphics[width=\linewidth]{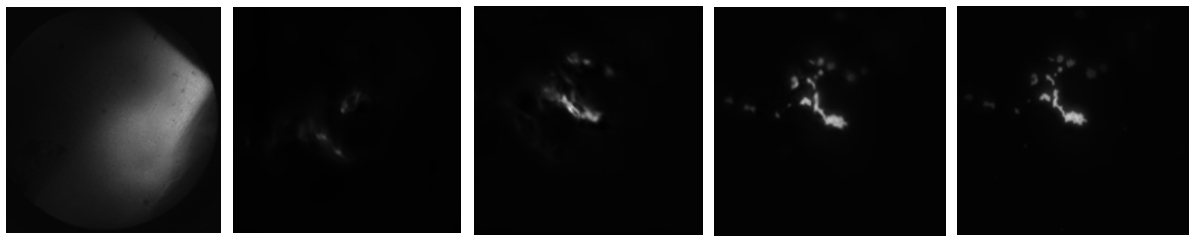} 
\caption{\small Example of optical reconstruction with different loss functions with SSIM and PSNR(dB) value. From left to right: $\bf y$, $\ell_2$$\{0.137, 48.756\}$, $\ell_2$ + $\ell_1$$\{0.251, 67.693\}$, $\ell_2 +$ SSIM $\{0.798, 73.012\}$, and $\bf x$.}
\label{exp:diff_loss}
\end{figure}

In this section, we investigate how different loss functions affect the reconstruction performance of our model. Specifically, we compare three configurations: pure $\ell_2$ loss, a combination of $\ell_2 + \ell_1$ losses, and a combined $\ell_2 + \text{SSIM}$ loss. The $\ell_2$ loss emphasizes pixel-wise accuracy, the $\ell_1$ loss encourages sparsity and robustness to outliers, whereas the SSIM loss focuses on preserving structural similarity, which is critical for perceptual quality. Table~\ref{tab:loss_comparison} and Figure~\ref{exp:diff_loss} show that models trained with the combination loss function $\ell_2$ + SSIM yield the best objective and subjective performance \citep{zhao2016loss}. 

\begin{table}[htb]
\centering
\caption{Comparison of model reconstruction results when trained with different loss functions}
\label{tab:loss_comparison}
\resizebox{\textwidth}{!}{
\begin{tabular}{@{}lccccc@{}}
\toprule
\textbf{Loss Function} & \textbf{MSE}  & \textbf{MAE}  & \textbf{PSNR (dB)}  & \textbf{SSIM}  & \textbf{FDR ($\times$10$^{-2}$)} \\
\midrule
\textbf{$\ell_2$}             & 0.111 $\pm$ 0.25        & 0.318 $\pm$ 0.073         & 49.69 $\pm$ 3.01              & 0.101 $\pm$ 0.0148           & 1.057 $\pm$ 0.64   \\
\textbf{$\ell_2 + \ell_1$}       & 0.0101 $\pm$ 0.18       & 0.0797 $\pm$ 0.092        & 67.083 $\pm$ 2.15             & 0.243 $\pm$ 0.053           & 1.055 $\pm$ 0.41   \\
\textbf{$\ell_2$} + SSIM & \textbf{\textcolor{red}{0.00431 $\pm$ 0.0013}} & \textbf{\textcolor{red}{0.0253 $\pm$ 0.0073}} & \textbf{\textcolor{red}{71.992 $\pm$ 1.94}} & \textbf{\textcolor{red}{0.814 $\pm$ 0.069}} & \textbf{\textcolor{red}{0.901 $\pm$ 0.22}} \\
\bottomrule
\end{tabular}
}
\end{table}




\subsection{Ablation Study: The Role of Skip Connections}

We investigate here the importance of skip connections in the TRUST architecture by analyzing how their removal affects reconstruction performance. Skip connections 
enable the direct transfer of low-level spatial features from the encoder to the decoder \citep{mao2016image, he2016deep}. These connections play a vital role during upsampling, allowing the model to recover refined supports and details that may otherwise be lost in the bottleneck layer.

\begin{figure}[!htb]
\centering
\includegraphics[width=\linewidth]{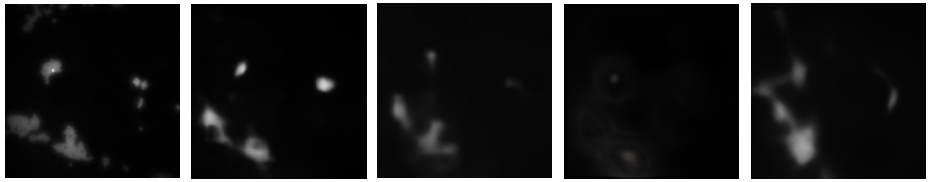} 
\caption{\small Different skip-connection reconstruction results with SSIM and PSNR(dB) value. From left to right: target $\bf x$, TRUST $\{0.862, 72.744\}$, TRUST mv skip1$\{0.610, 71.662\}$, TRUST mv skip1 \& skip2$\{0.304, 67.832\}$, TRUST with no skip$\{0.654, 69.512\}$}
\label{exp:skip}
\end{figure}

To quantify their impact, we conduct a series of ablation experiments by systematically disabling skip connections at different stages of the TRUST network. As shown in Table~\ref{tab:skip_comparison} and visualized in Figure~\ref{exp:skip}, removing even a single skip connection results in a noticeable drop in performance across all evaluation metrics. The degradation is particularly pronounced in high-frequency regions and structural boundaries, where spatial detail is most critical. These findings reaffirm the importance of skip connections in preserving spatial fidelity and demonstrate their indispensable role in enabling high-quality image reconstruction within the TRUST framework.


\begin{table}[h]
\centering
\caption{Impact of Skip Connections on Reconstruction Performance}
\label{tab:skip_comparison}
\resizebox{\textwidth}{!}{
\begin{tabular}{@{}lccccc@{}}
\toprule
\textbf{Configuration}    & \textbf{MSE}  & \textbf{MAE}  & \textbf{PSNR (dB)}  & \textbf{SSIM}  & \textbf{FDR ($\times 10^{-2}$)} \\
\midrule
\textbf{TRUST}                 & \textbf{\textcolor{red}{0.00431 $\pm$ 0.0013}} & \textbf{\textcolor{red}{0.0253 $\pm$ 0.0073}} & \textbf{\textcolor{red}{71.992 $\pm$ 1.94}} & \textbf{\textcolor{red}{0.814 $\pm$ 0.069}} & \textbf{\textcolor{red}{0.901 $\pm$ 0.22}} \\
\textbf{TRUST mv skip1}        & 0.00441 $\pm$ 0.0027  & 0.0280 $\pm$ 0.011  & 71.082 $\pm$ 1.91  & 0.774 $\pm$ 0.065  & 1.223 $\pm$ 0.28 \\
\textbf{TRUST mv skip1 \& skip2} & 0.00681 $\pm$ 0.0046 & 0.0468 $\pm$ 0.023  & 70.156 $\pm$ 2.18  & 0.610 $\pm$ 0.1322  & 3.034 $\pm$ 0.64 \\
\textbf{TRUST no skip}          & 0.00540 $\pm$ 0.0021 & 0.0314 $\pm$ 0.011 & 70.990 $\pm$ 1.80  & 0.746 $\pm$ 0.062 & 1.640 $\pm$ 0.47 \\
\bottomrule
\end{tabular}
}
\end{table}

\begin{table}[h]
\caption{How pretrained attention impact reconstruction results}
\label{tab:pre}
\resizebox{\textwidth}{!}{
\begin{tabular}{@{}lccccc@{}}
\toprule
\textbf{Method} & \textbf{MSE} & \textbf{MAE} & \textbf{PSNR (dB)} & \textbf{SSIM} & \textbf{FDR($\times$10$^{-2}$)} \\
\midrule
\textbf{TRUST without Pretrained ViT}  & 0.00601 $\pm$ 0.0034      & 0.0341 $\pm$ 0.014       & 70.583 $\pm$ 1.81              & 0.697 $\pm$ 0.072        & 2.093 $\pm$ 0.19       \\
\textbf{TRUST with Pretrained ViT}  & \textbf{\textcolor{red}{0.00431 $\pm$ 0.0013}} & \textbf{\textcolor{red}{0.0253 $\pm$ 0.0073}} & \textbf{\textcolor{red}{71.992 $\pm$ 1.94}} & \textbf{\textcolor{red}{0.814 $\pm$ 0.069}} & \textbf{\textcolor{red}{0.901 $\pm$ 0.22}} \\
\bottomrule
\end{tabular}
}
\end{table}


\subsection{Ablation Study: Pretraining vs. Training from Scratch}


\begin{figure}[!htb]
\centering
\includegraphics[width=\linewidth]{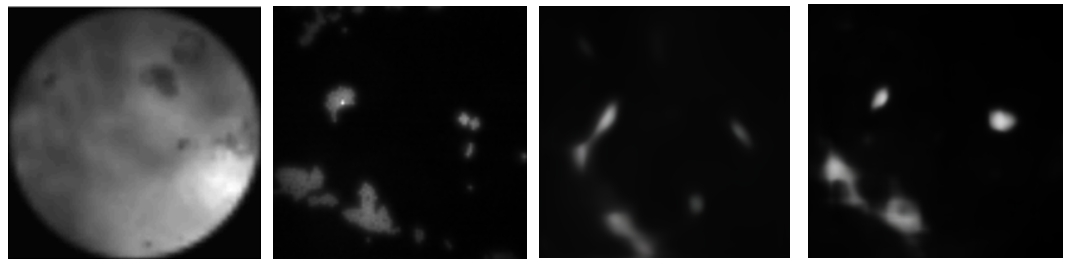} 
\caption{Pretrained-vs-Not reconstruction results with SSIM and PSNR(dB) value. From left to right: target, TRUST without pretraining $\{0.606, 71.342\}$, TRUST with pretraining $\{0.862, 72.744\}$}.
\label{exp:pre_nopre}
\end{figure}
\vspace{-0.135in}

In this ablation study, we evaluate the effect of pretraining in the Vision Transformer (ViT) encoder on the performance of the TRUST architecture. Specifically, we compare two configurations: one using a pretrained ViT (initialized with weights from the ‘google/vit-base-patch16-224’ model) and another where the attention encoder is trained from scratch on the target dataset.

Leveraging a pretrained ViT allows the model to start from a strong feature representation that captures generalized and discriminative patterns, even when fine-tuned on relatively small domain-specific datasets \citep{chen2019med3d}. As reported in Table~\ref{tab:pre} and visualized in Figure~\ref{exp:pre_nopre}, ViT pretraining significantly enhances reconstruction performance across all evaluation metrics. These results highlight the effectiveness of transfer learning in boosting the feature extraction capacity of the attention module. By starting with a rich, pretrained representation, the model converges faster and produces reconstructions that are not only quantitatively superior but also perceptually more accurate.


\section{Conclusion and Future Work}
In this paper, we introduced TRUST, a hybrid architecture that integrates a pretrained Vision Transformer (ViT) encoder with a U-Net decoder for high-quality sparse image reconstruction. Experimental results show that TRUST consistently outperforms both classical and deep learning baselines, achieving superior performance across standard metrics, including PSNR, SSIM, MSE, MAE, and FDR, while significantly reducing hallucination artifacts.

TRUST’s effectiveness is attributed to its key architectural components: {\it (i)} a ViT-based attention encoder that captures global dependencies early in the pipeline; {\it (ii)} skip connections that enable multi-scale feature fusion; and {\it (iii)} a hierarchical decoder that refines coarse global representations into high-resolution image details. 
Despite its advantages, TRUST introduces additional computational overhead due to its reliance on a pretrained transformer backbone, resulting in $2-3\times$ higher inference time compared to U-Net under equivalent hardware conditions. Also, while this study focuses on sparse optical image recovery, the underlying design principles of TRUST -- attention-guided global context modeling and hierarchical multiresolution decoding -- are broadly applicable \citep{touvron2021training}. 
Future work will explore TRUST extensions to various signal processing tasks while also addressing the model’s computational complexity to improve efficiency and scalability \citep{mehta2021mobilevit}.

\bibliographystyle{unsrt}

\newpage
\appendix
\section*{Appendix}

\section{Error Bound for the Attention Mechanism}
We assume that we have two tokens $\bf x$ and $\bf y$, which are related via the linear constraint $\bf y = A x$. In practice, most of the time we have some additional prior knowledge on the operator $\bf A$ (after all, we typically design an appropriate $\bf A$ for the application at hand) such as:
\begin{itemize}
\item $\bf A$ is orthonormal square matrix; or
\item $\bf A$ is tall matrix with orthonormal columns; or
\item $\bf A$ is fat matrix satisfying the Restricted Isometry Property (RIP).
\end{itemize}

The attention mechanism is formulated as
\begin{equation}
    \text{Attention}({\bf Q}, {\bf K}, {\bf V}) = \text{softmax}\left(\frac{{\bf QK}^T}{\sqrt{d_k}}\right) {\bf V}
    \label{attention2}
\end{equation}
Performing self attention on $\bf y$ yields the following:
\begin{equation}
    \text{Attention}({\bf y}) = \text{softmax}\left(\frac{{\bf y}^T {\bf y}}{\sqrt{d_k}}\right) {\bf V} = \text{softmax}\left(\frac{{\bf x}^T {\bf A}^T {\bf A x}}{\sqrt{d_k}}\right) {\bf V}.
    \label{attention3}
\end{equation}

When $\bf A$ has orthonormal columns, it is clear that attention above yields the same value in either $\bf x$ or $\bf y$ domain. In compressed sensing applications, $\bf A$ is most likely fat and the orthonormal property of its columns breaks down. In this case, we need to rely on the RIP of $\bf A$ as follows:  
let \( {\bf A} \in \mathbb{R}^{m \times n} \) be a matrix satisfying the Restricted Isometry Property (RIP) of order \( 2k \) with constant \( \delta_{2k} \in (0, 1) \). That is, for all \( 2k \)-sparse vectors \( {\bf z} \in \mathbb{R}^n \), we have
\[
(1 - \delta_{2k}) \| {\bf z} \|_2^2 \leq \| {\bf A z} \|_2^2 \leq (1 + \delta_{2k}) \| {\bf z} \|_2^2.
\]

Let \( {\bf x, x'} \in \mathbb{R}^n \) be two normalized vectors with supports of size at most \( k \), i.e., both are \( k \)-sparse and $\| {\bf x} \|_2^2 = \| {\bf x'} \|_2^2 = 1$. Then, their sum or difference support together has size at most \( 2k \). In other words, \( {\bf x} + {\bf x'} \) and \( {\bf x} - {\bf x'} \) are \( 2k \)-sparse. We aim to bound the following difference between the original and transformed inner product:
\[
\left| {\bf x}^\top {\bf A}^\top {\bf A x'} - {\bf x}^\top {\bf x'} \right|.
\]

The polarization identity combined with the RIP condition yields:
\begin{align*}
\| {\bf A}( {\bf x} + {\bf x'})\|_2^2 &= \| {\bf A x} \|_2^2 + 2 {\bf x}^\top {\bf A}^\top {\bf A x'} + \| {\bf A x'} \|_2^2, \\
\| {\bf A} ({\bf x} - {\bf x'})\|_2^2 &= \| {\bf A x} \|_2^2 - 2 {\bf x}^\top A^\top {\bf A x'} + \| {\bf A x'} \|_2^2.
\end{align*}

Subtracting these two identities gives:
\[
\| {\bf A} ({\bf x} + {\bf x'})\|_2^2 - \|{\bf A} ({\bf x} - {\bf x'})\|_2^2 = 4 {\bf x}^\top {\bf A}^\top {\bf A x'}.
\]

Similarly, if $\bf A$ is the identity matrix, we have:
\[
\| {\bf x} + {\bf x'}\|_2^2 - \| {\bf x} - {\bf x'}\|_2^2 = 4 {\bf x}^\top {\bf x'}.
\]

Imposing RIP on \( {\bf x + x'} \) and \( {\bf x - x'} \) produces 
\[
\left| \| {\bf A} ( {\bf x + x'} )\|_2^2 - \| {\bf x + x'} \|_2^2 \right| \leq \delta_{2k} \| {\bf x + x'} \|_2^2,
\]
\[
\left| \| {\bf A} ({\bf x - x'})\|_2^2 - \| {\bf x - x'} \|_2^2 \right| \leq \delta_{2k} \|{\bf x - x'} \|_2^2.
\]

Combining the two and applying the triangle inequality, we can finally obtain the following bound:
\begin{align*}
\left| {\bf x}^\top {\bf A}^\top {\bf A x'} - {\bf x}^\top {\bf x'} \right|
&= \frac{1}{4} \left| \left( \| {\bf A} ( {\bf x + x'} )\|_2^2 - \| {\bf A} ({\bf x - x'})\|_2^2 \right) - \left( \|{\bf x + x'}\|_2^2 - \|{\bf x - x'}\|_2^2 \right) \right| \\
&\leq \frac{1}{4} \left( \left| \|{\bf A} ({\bf x + x'})\|_2^2 - \| {\bf x + x'} \|_2^2 \right| + \left| \|{\bf A} ({\bf x - x'})\|_2^2 - \| {\bf x - x'} \|_2^2 \right| \right) \\
&\leq \frac{\delta_{2k}}{4} \left( \| {\bf x + x'} \|_2^2 + \| {\bf x - x'} \|_2^2 \right) \\
&= \frac{\delta_{2k}}{4} \left( 2\| {\bf x} \|_2^2 + 2\| {\bf x'} \|_2^2 \right) \\
&= \frac{\delta_{2k}}{2} \left( \| {\bf x} \|_2^2 + \| {\bf x'} \|_2^2 \right)\\
&= \frac{\delta_{2k}}{2} \left( 1 + 1 \right)\\
&= \delta_{2k}.
\end{align*}

\begin{figure}[!htb]
\centering
\includegraphics[width=\linewidth]{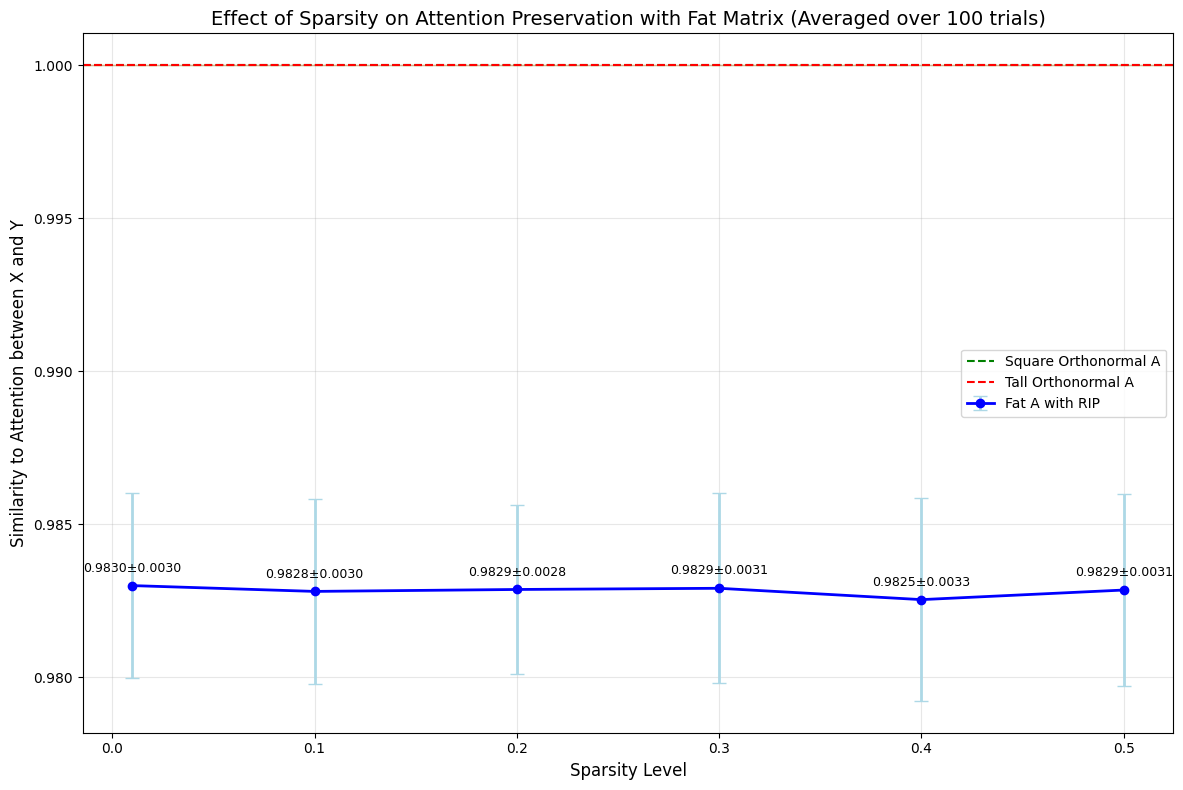} 
\caption{Simulation of similarity between attention on $\bf x$ and $\bf y = A x$} for various sensing matrices $\bf A$'s.
\label{appd: attention_similarity}
\end{figure}

Figure~\ref{appd: attention_similarity} illustrates the average effect of sparsity and fat random Gaussian matrices on attention/similarity averaged over 100 totally random trials. As expected, $\bf A$'s with orthonormal columns yield exactly the same attention. On the other hand, we confirm that we are still able to obtain close approximation of the attention level with fat random Gaussian sensing matrices $\bf A$'s.     


\newpage

\section{Evaluation Metrics}

To evaluate the reconstruction quality of our models, we employ both standard image similarity metrics and a custom hallucination-aware metric:

\paragraph{Root Mean Squared Error (RMSE).}
RMSE measures the square root of the average squared differences between predicted and ground truth pixel values:
\[
\text{RMSE} = \sqrt{\frac{1}{N} \sum_{i=1}^{N} (x_i - \hat{x}_i)^2},
\]
where \( x_i \) and \( \hat{x}_i \) are the ground truth and predicted pixel values, respectively.

\paragraph{Peak Signal-to-Noise Ratio (PSNR).}
PSNR quantifies the reconstruction fidelity relative to the maximum pixel intensity:
\[
\text{PSNR} = 20 \cdot \log_{10} \left( \frac{\text{MAX}}{\text{RMSE}} \right),
\]
where \(\text{MAX}\) is the maximum possible pixel value (assumed to be 1.0 after normalization).

\paragraph{Structural Similarity Index Measure (SSIM).}
SSIM evaluates perceptual image similarity by comparing local patterns of luminance, contrast, and structure. The score ranges from \(-1\) to 1, with 1 indicating perfect structural alignment.


\paragraph{False Positive Region Score (FPR).}
We define a hallucination-sensitive metric called the False Positive Region (FPR) score to quantify spurious regions generated by the model. A pixel is considered hallucinated if it satisfies:
\[
x_{\text{hat}} > t_{\text{high}} \quad \text{and} \quad x_{\text{true}} \leq t_{\text{low}},
\]
The FPR score is computed as the fraction of hallucinated pixels over the entire image:
\[
\text{FPR} = \frac{\left| \{ i : x_{\text{hat}, i} > t_{\text{high}} \ \wedge\ x_{\text{true}, i} \leq t_{\text{low}} \} \right|}{N}.
\]

\section{Extended Sparse Recovery Results}

All the models listed below were trained with approximately same hyper-parameters as specified in the paper, and the stop condition is when reaching the nearly same loss values. This setup ensures a fair comparison under similar consistent conditions.

\subsection{Extended Results on Sparse Recovery of Optics Data}

In this section, we present a more comprehensive comparison of model performance on sparse recovery tasks using the optical imaging dataset.

\begin{figure}[!htb]
\centering
\includegraphics[width=\linewidth]{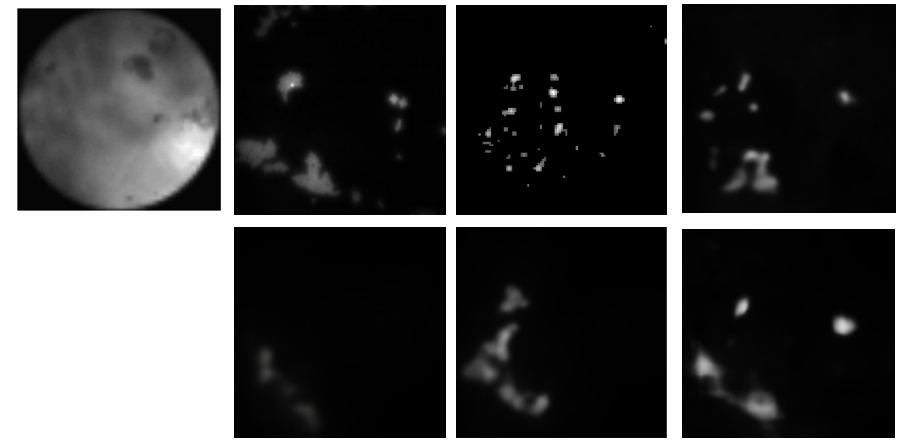} 
\caption{\small Example of reconstruction results with corresponding SSIM and PSNR values. Top row, from left to right: response $\bf y$, target $\bf x$, OMP $\{0.301, 68.723dB\}$, and U-Net $\{0.779, 71.691dB\}$. Bottom row, from left to right: TransUnet $\{0.672, 67.236dB\}$, Restormer $\{0.752, 71.762dB\}$, and TRUST $\{0.862, 72.744dB\}$}.
\label{appd: optics_1}
\end{figure}

\begin{figure}[!htb]
\centering
\includegraphics[width=\linewidth]{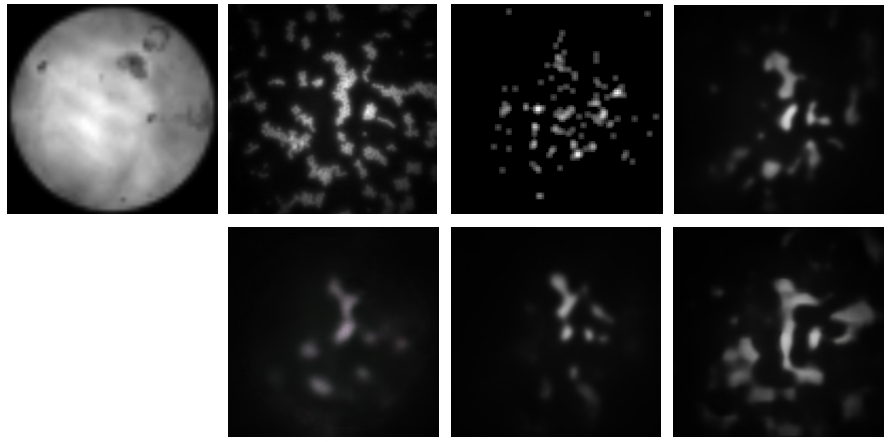} 
\caption{\small Example of reconstruction results with corresponding SSIM and PSNR values. Top row, from left to right: response $\bf y$, target $\bf x$, OMP $\{0.325, 63.071dB\}$, and U-Net $\{0.636, 66.712dB\}$. Bottom row, from left to right: TransUnet $\{0.553, 66.351dB\}$, Restormer $\{0.625, 66.583dB\}$, and TRUST $\{0.671, 68.276dB\}$}.
\label{appd: optics_2}
\end{figure}

\begin{figure}[!htb]
\centering
\includegraphics[width=\linewidth]{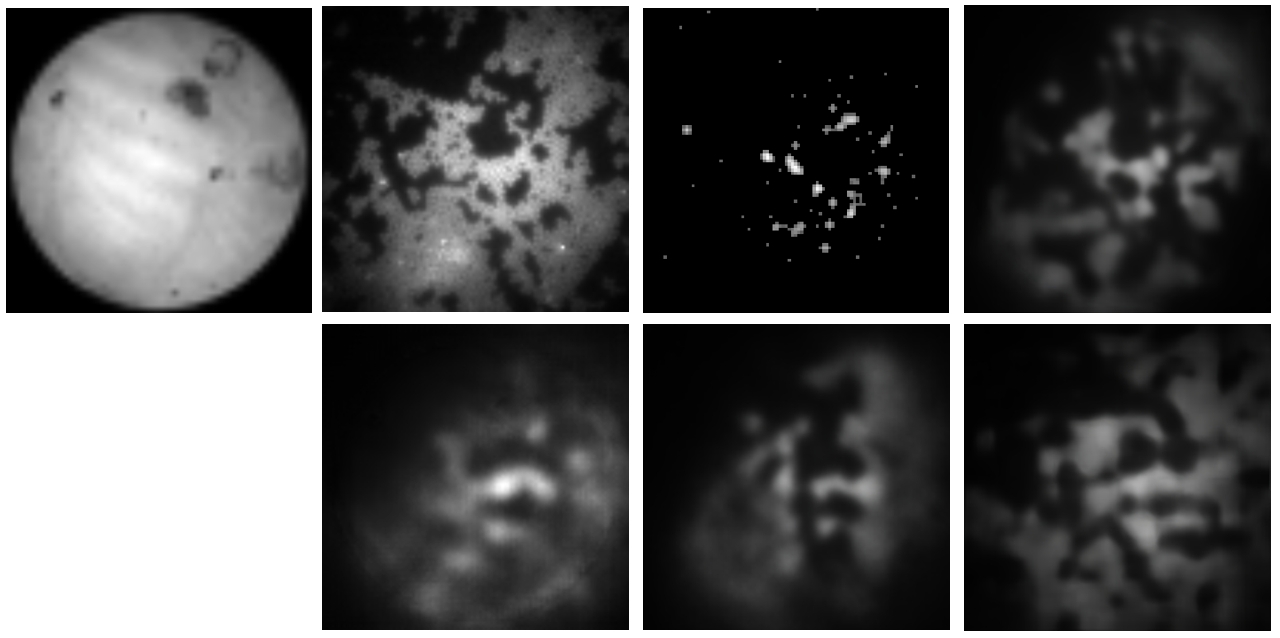} 
\caption{\small Example of reconstruction results with corresponding SSIM and PSNR values. Top row, from left to right: response $\bf y$, target $\bf x$, OMP $\{0.244, 58.232dB\}$, and U-Net $\{0.513, 62.105dB\}$. Bottom row, from left to right: TransUnet $\{0.409, 61.812dB\}$, Restormer $\{0.542, 62.503dB\}$, and TRUST $\{0.592, 63.427dB\}$}.
\label{appd: optics_3}
\end{figure}

Figures~\ref{appd: optics_1}, \ref{appd: optics_2}, and \ref{appd: optics_3} illustrate qualitative reconstruction results across various models, while the quantitative metrics are summarized in Table~\ref{tab:optics_recon_appendix}. The data clearly show that TRUST consistently outperforms all competing neural network architectures, achieving superior reconstruction fidelity across all evaluation criteria.

As expected, traditional sparse recovery methods deliver the weakest performance, producing reconstructions with significant artifacts and loss of structural detail. Among deep learning models, the fully transformer-based Restormer yields competitive results but exhibits a consistent tendency to under-predict fine-scale features, leading to a higher missing probability error. This suggests that despite its strong global modeling capabilities, Restormer may struggle to capture the fine-grained spatial details necessary for precise optical reconstruction.

These results reinforce the advantage of TRUST’s hybrid architecture, which leverages both global attention mechanisms and localized multi-scale refinement to achieve accurate and perceptually faithful image recovery.

\begin{table}[tbh]
\centering
\small
\caption{Average recovery performance on the optics dataset: mean $\pm$ standard deviation}
\label{tab:optics_recon_appendix}
\begin{tabular}{@{}lccccc@{}}
\toprule
\textbf{Method}       & \textbf{MSE}               & \textbf{MAE}              & \textbf{PSNR (dB)}             & \textbf{SSIM}             & \textbf{FDR ($\times$10$^{-2}$)} \\
\midrule
\textbf{OMP}          & 0.0111 $\pm$ 0.0032        & 0.0435 $\pm$ 0.0062       & 68.04 $\pm$ 2.03               & 0.279 $\pm$ 0.035        & 5.30 $\pm$ 1.03               \\
\textbf{U-Net}        & 0.00451 $\pm$ 0.0022       & 0.0398 $\pm$ 0.012        & 70.76 $\pm$ 2.00               & 0.772 $\pm$0.053          & 1.14 $\pm$ 0.16              \\
\textbf{TransUNet}    & 0.00911 $\pm$ 0.0040       & 0.0440 $\pm$ 0.012        & 69.84 $\pm$ 1.92               & 0.636 $\pm$0.091          & 2.61 $\pm$ 3.1              \\
\textbf{Restormer}    & 0.00823 $\pm$ 0.0041       & 0.0405 $\pm$ 0.013        & 70.48 $\pm$ 2.13               & 0.715 $\pm$0.056          & 0.907 $\pm$ 0.36              \\
\textbf{TRUST}        & \textbf{\textcolor{red}{0.00431 $\pm$ 0.0013}} & \textbf{\textcolor{red}{0.0253 $\pm$ 0.0073}} & \textbf{\textcolor{red}{71.992 $\pm$ 1.94}} & \textbf{\textcolor{red}{0.814 $\pm$ 0.069}} & \textbf{\textcolor{red}{0.901 $\pm$ 0.22}} \\
\bottomrule
\end{tabular}
\end{table}

\newpage
\subsection{Extended Results on Sparse Recovery of FastMRI Data}


This section presents an extended comparison of sparse recovery performance on the FastMRI dataset across four deep neural network architectures. 

\begin{figure}[!htb]
\centering
\includegraphics[width=\linewidth]{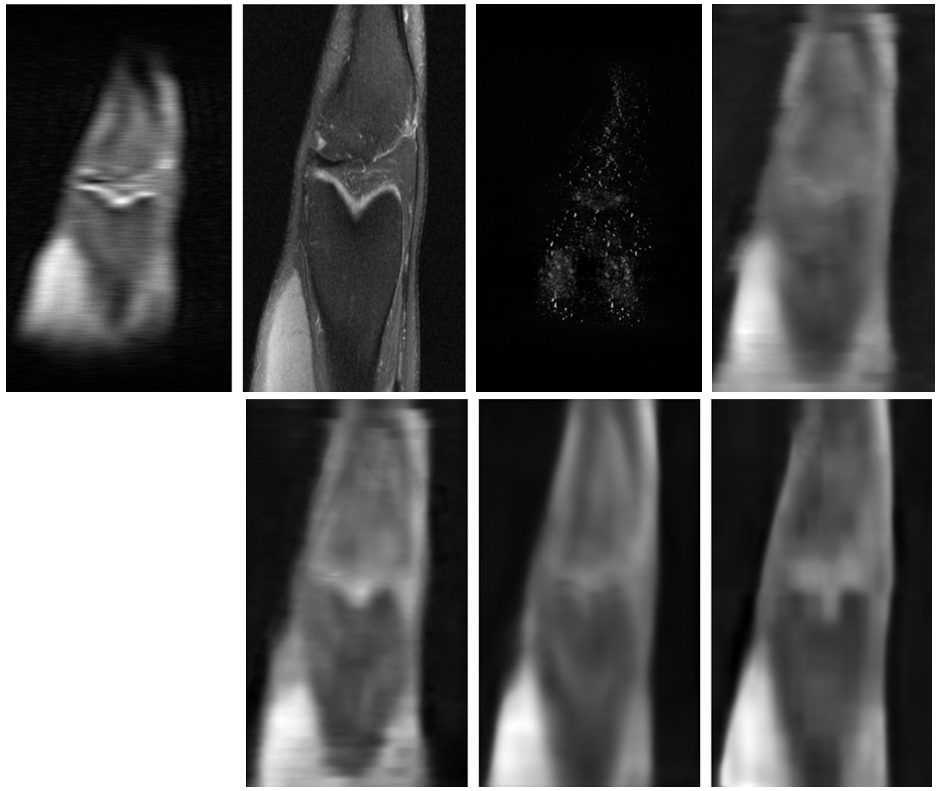} 
\caption{\small Example of reconstruction results with corresponding SSIM and PSNR values. Top row, from left to right: undersampled input $\bf y$, target $\bf x$, OMP $\{0.173, 15.682dB\}$, U-Net $\{0.610, 21.623dB\}$. Bottom row, from left to right: TransUnet $\{0.614, 21.956dB\}$, Restormer $\{0.623, 22.631dB\}$, and TRUST $\{0.629, 22.893dB\}$}.
\label{appd: mri_1}
\end{figure}

\begin{figure}[!htb]
\centering
\includegraphics[width=\linewidth]{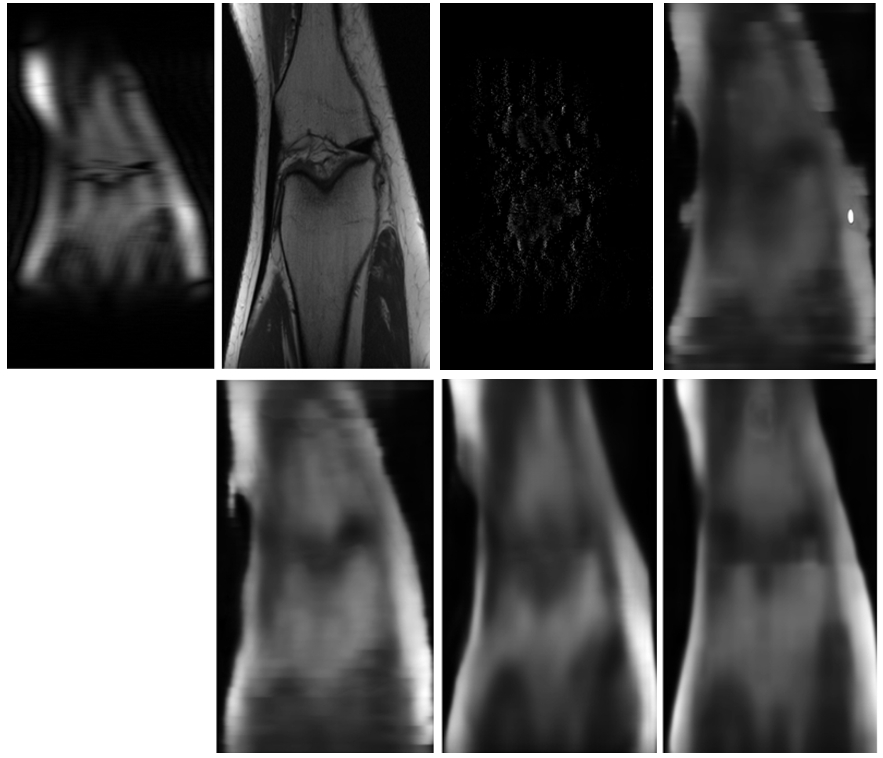} 
\caption{\small Example of reconstruction results with corresponding SSIM and PSNR values. Top row, from left to right: undersampled input $\bf y$, target $\bf x$, OMP $\{0.2430, 12.812dB\}$, U-Net $\{0.612, 18.844dB\}$. Bottom row, from left to right:: TransUnet $\{0.635, 19.593dB\}$, Restormer $\{0.636, 20.271dB\}$, and TRUST $\{0.687, 21.593dB\}$}.
\label{appd: mri_2}
\end{figure}

\begin{figure}[!htb]
\centering
\includegraphics[width=\linewidth]{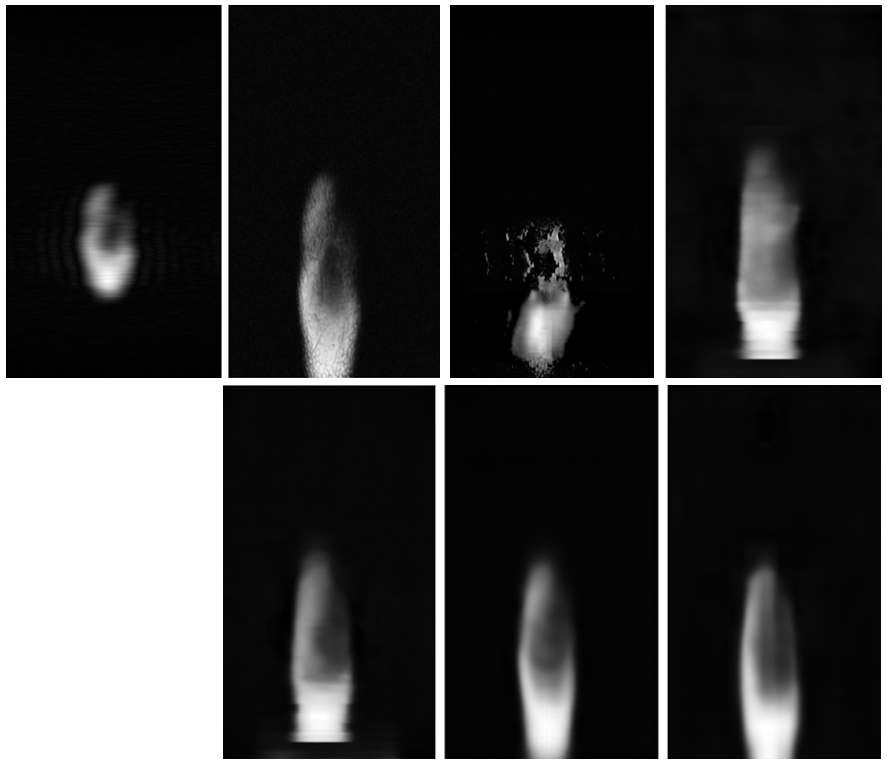} 
\caption{\small Example of reconstruction results with corresponding SSIM and PSNR values. Top row, from left to right: undersampled input $\bf y$, target $\bf x$, OMP $\{0.5230, 19.083dB\}$, U-Net $\{0.586, 21.693dB\}$, TransUnet $\{0.871, 22.631dB\}$, Restormer $\{0.877, 26.568dB\}$, and TRUST $\{0.889, 30.602dB\}$}.
\label{appd: mri_3}
\end{figure}

Figures~\ref{appd: mri_1}, \ref{appd: mri_2}, and \ref{appd: mri_3} showcase representative examples of MRI image reconstruction under typical k-space undersampling scenarios. The corresponding quantitative results are summarized in Table~\ref{tab:MRI2}, which reports the mean and standard deviation of recovery performance across approximately 3,000 test images.

Consistent with earlier findings, our proposed hybrid model TRUST outperforms all competing approaches in both objective and subjective measures. It achieves higher reconstruction quality as measured by standard metrics and produces visibly more faithful image details -- highlighting the effectiveness of TRUST’s architecture in capturing both global structure and fine-grained spatial information in complex medical imaging tasks. 





\begin{table}[htb]
\centering
\caption{Average recovery performance on the FastMRI dataset: mean $\pm$ standard of deviation}
\label{tab:MRI2}
\begin{tabular}{@{}lccccc@{}}
\toprule
\textbf{Method} & \textbf{MSE} & \textbf{MAE} & \textbf{PSNR (dB)} & \textbf{SSIM} & \textbf{FDR($\times$10$^{-2}$)} \\
\midrule
\textbf{OMP}  & 0.109 $\pm$ 0.543      & 0.138 $\pm$ 0.0923      & 14.37 $\pm$ 4.34            & 0.145 $\pm$ 0.0395         & 6.26 $\pm$ 3.22  \\
\textbf{U-Net}  & 0.0861 $\pm$ 0.0246      & 0.0506 $\pm$ 0.0174      & 21.70 $\pm$ 2.74            & 0.668 $\pm$ 0.0900         & 4.26 $\pm$ 4.99       \\
\textbf{TransUNet}  & 0.0703 $\pm$ 0.0208      & 0.0396 $\pm$ 0.0178      & 21.07 $\pm$ 2.34            & 0.6553 $\pm$ 0.0863         & 5.93 $\pm$ 6.21       \\
\textbf{Restormer}  & 0.0692 $\pm$ 0.0227      & 0.0411 $\pm$ 0.0160       & 23.72 $\pm$ 3.15              & 0.698 $\pm$ 0.0953        & 2.97 $\pm$ 4.74       \\
\textbf{TRUST}  & \textbf{\textcolor{red}{0.0613 $\pm$ 0.0220}} & \textbf{\textcolor{red}{0.0353 $\pm$ 0.0133}} & \textbf{\textcolor{red}{24.81 $\pm$ 3.13}} & \textbf{\textcolor{red}{0.717 $\pm$ 0.0851}} & \textbf{\textcolor{red}{2.78 $\pm$ 4.33}} \\
\bottomrule
\end{tabular}
\end{table}

\section{Model and Computational Complexity Comparison}

In this section, we provide a brief supplemental comparison of the model complexity and computational efficiency of four competing deep neural network architectures: TRUST, TransUNet, Restormer, and U-Net.

While the TRUST model demonstrates strong performance across all tasks presented in previous sections, its reliance on the ViT-base backbone results in a relatively high parameter count of approximately 9 million, which is comparable to TransUNet. In contrast, Restormer maintains a smaller footprint at 3 million parameters, and U-Net remains the most lightweight, with only 2 million parameters.

In terms of training complexity, TRUST, TransUNet, and U-Net exhibit similarly efficient training behavior. Using the modest hardware configuration described earlier, each model completes 50 epochs of training in approximately 24 hours. By comparison, Restormer is significantly more computationally demanding: under the same conditions, it progresses through only 8 epochs in a 24-hour period, highlighting its heavier training requirements.

For inference speed, U-Net is the fastest, generating images in roughly 0.006 seconds per frame, owing to its simple architecture. TRUST and TransUNet take slightly longer, averaging 0.013 seconds per image, while Restormer, with its deeper and more complex architecture, requires approximately 0.06 seconds per image.

Despite these computational trade-offs, we would like to make the following final note: the TRUST model has not yet been fully optimized. Our long-term goal is to deploy TRUST for real-time image reconstruction directly from optical system measurements. The current results suggest that reducing the computational load of the ViT-based encoder is a promising direction. In future work, we aim to explore more lightweight, task-specific attention modules that can serve as efficient substitutes for the full transformer block -- potentially preserving or improving performance while significantly decreasing computational overhead.









\end{document}